\documentclass[aps,prl,groupedaddress,showpacs,superscriptaddress,twocolumn]{revtex4}

\usepackage{graphicx}
\usepackage{dcolumn}
\usepackage{bm}
\begin{document}

\title{Non-linear transport and heat dissipation in metallic carbon nanotubes}


\author{Marcelo A. Kuroda}
\affiliation{Beckman Institute and Department of Physics, University of Illinois at Urbana-Champaign, Illinois 61801}
\author{Jean-Pierre Leburton}
\affiliation{Beckman Institute and Department of Electrical and Computer Engineering, University of Illinois at Urbana-Champaign, Illinois 61801}


\date{\today}

\begin{abstract}
 We show that the local temperature dependence of
thermalized electron and phonon populations along metallic carbon nanotubes
is the main reason behind this non-linear transport characteristics
in the high bias regime. Our model that considers optical and zone boundary 
phonon emission as well as absorption by charge carriers is based 
on the solution of the Boltzmann transport equation that assumes a 
local temperature along
the nanotube, determined self-consistently with the heat transport
equation. By using realistic transport parameters, our results not 
only reproduce experimental data for
electronic transport, but also provide a coherent interpretation of
thermal breakdown under electric stress. In particular, electron and phonon
thermalization prohibits ballistic transport in short nanotubes.
\end{abstract}


\maketitle

Carbon nanotubes (CN) are one-dimensional (1D) nanostructures that
have stimulated broad research interest because of their unique
electrical versatility into semiconductors and metals, depending
of their chirality \cite{saito_dres}. From a technological
viewpoint, their remarkable electrical and mechanical properties
make them promising materials for applications in high performance
nanoscale electronic and mechanical devices \cite{mceuen,avouris}.
Among these properties, the interrelation between electronic and
thermal transport in these quasi 1D structures is particularly interesting.
Early experiments on non-linear transport in metallic single
walled nanotubes (m-SWNTs) using low resistance contacts revealed
current saturation at the 25$\mu$A level, which was attributed to
the onset of electron backscattering by high energy optical (OP)
and zone boundary (ZB) phonons in the high bias regime \cite{yao}.
More recently, series of independent high-field transport
measurements on various length m-SWNTs demonstrated the
absence of current saturation with current levels over 60$\mu$A in
short samples ($\lesssim$55nm), which was interpreted as ballistic
transport along the CN \cite{javey,park}. Among the findings was
also the observation of thermal breakdown and burning under high
electric stress. In the mean time, the electrical conductance of
multi-walled nanotubes (MWNTs) under high bias has shown step-like
decrease caused by the successive burning of the CN outer shells
\cite{collins1,collins2}. In these experiments CNs burn
unexpectedly at mid-length under stress even on a substrate and
on the presence of a back gate in a field effect device geometry
\cite{muller}. Despite various attempts to model these systems
\cite{yao,javey,park}, up to now no coherent interpretation has
emerged that reconciliates heat dissipation with electronic
transport and describes thermal effects in m-CNs under electric
stress.

In this letter, we show that the non-linear characteristics of
metallic CNs find their origin in the non-homogeneous Joule heating
along the nanotube, which is caused by the thermalized
distribution of electrons scattered by high energy phonons, even
in short m-SWNT. We specifically show that Joule heating is
maximum at CN mid-length and, owing to the 1D nature of the
structure, increases drastically  with the CN length, resulting in
thermal breakdown at lower bias than in shorter CNs. Our model is
based on the Boltzmann transport equation with OP and ZB phonon
scattering and solved self-consistently with the heat transfer
equation, providing a coherent interpretation of electric and
thermal transport in m-SWNTs in agreement with experimental data
\cite{javey, park}. In particular we show that the high current
level in short CNs is not due to ballistic transport but to
reduced Joule heating.

\begin{figure}[htbp]
  \leavevmode \centering
    \includegraphics[width=3in]{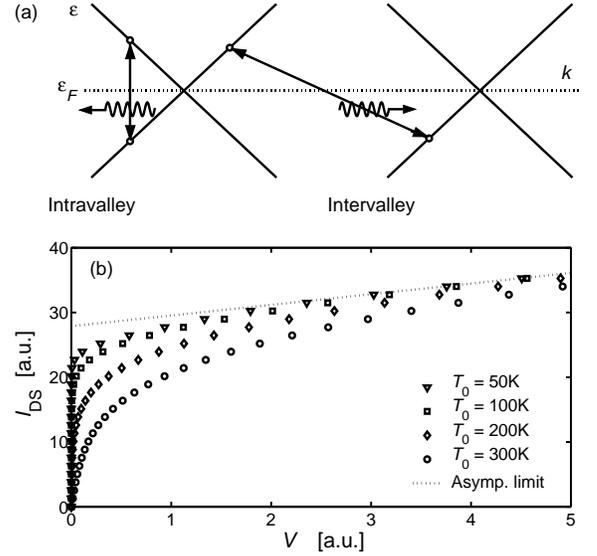}
    \vspace{-0.15in}
  \caption{\label{iv_const_diag} (a) Scattering processes considered in
the calculations. The intravalley (left) and intervalley (right)
transitions with emission and absorption of optical phonons are
included. (b) $IV$ characteristics for different constant
temperatures along the tube. Dashed line: asymptotic behavior
for all temperatures in the high bias regime.}
\end{figure}

We use the linear dispersion relation of electronic states
around the Fermi level ($\epsilon(k)\! =\! \pm \hbar v_F k$) \cite{jishi},
 being $v_F$ the Fermi velocity. Thus,
the Boltzmann equation reads:
\begin{equation}
v_F \,\partial_x f^\alpha(\epsilon) + \frac{eF}{\hbar} \partial_k
f^\alpha(\epsilon) \label{boltzmann1} = C_{ph}^\alpha(I,T)
\label{boltzmann}
\end{equation}
Here $f^\alpha(\epsilon)$, $F$, $e$ and  $C_{ph}$ are the
distribution function for the $\alpha$ energy branch, the electric
field, the electron charge and the electron-phonon collision
integral, respectively. The index $\alpha$ denotes the energy
branches with positive (+) and negative (--) Fermi velocity in the
first (1) and second valleys (2) of the m-CN. In metallic systems
high electron density and strong inter-carrier scattering
thermalizes the electron distribution. We therefore assume that
the electron distribution function $f^\alpha\!(\epsilon)$ obeys
Fermi-Dirac statistics with a local electronic temperature
$T_{el}(x)$:
\begin{equation}
f^\alpha(\epsilon) = 1/(1+\exp((\epsilon -
\epsilon_F^\alpha)/k_BT_{el}(x))
\end{equation}
where $\epsilon_F^\alpha$ is the quasi Fermi level of
branch $\alpha$.
As a result, the collision integral $C^\alpha_{ph}$ also depends
on the position. We neglect acoustic phonons that are only
relevant in the low bias regime, and consider the contribution of
high density OP and ZB phonons \cite{saito,perebeinos}
that play a central role in energy dissipation in the high
bias regime. As illustrated in Fig.~\ref{iv_const_diag}(a),
the different processes (inter- and intra-branch)
considered in $C_{ph}$ include both the emission and absorption of
these phonons with energy ($\hbar \omega_{op}\! \approx\! 0.2\mbox{eV}$) much
larger than thermal fluctuations at room temperature. Each of these phonons contributes to the collision
integral as follow:
\begin{eqnarray}
C_{ph}^\alpha\!(\!I,\!T(x\!)\!)\!=\!\!\!\sum_{i,\beta}\!\!
\left\{\!\!\frac{R_e^i}{\pi}\right.\!\!\left[\!f^\beta\!(k\!)\!
\left(\!1\!\!-\!\!\!f^\alpha\!(k\!\!-\!\!q)\!\right)
\!-\!\!f^\alpha\!(k)\!\left(\!1\!-\!\!f^\beta\!(k\!\!-\!\!q\!)\!\right)\!\right]\!\!\!\nonumber\\
\!+\!\left.\frac{R^i_a}{\pi}\!\!\left[f^\beta\!(k)\!\left(\!1\!-\!f^\alpha\!(k\!+\!q)\right)
\!-\!f^\alpha\!(k)\!\! \left(\!1\!-\!f^\beta\!(k\!+\!q)\!\right)\!\right]\!\!\right\}\qquad\quad
\label{collint}
\end{eqnarray}
where the Greek letter index $\beta$ runs over the two branches
and two valleys, $i$ stands for OP and ZB phonons
and $R^i_e$ ($R^i_a$) is the phonon emission (absorption) rate.
Hence the first (second) two terms in Eq.~\ref{collint}
corresponds to processes involving the emission (absorption) of a
phonon limited by Pauli exclusion principle. For instance, the
first term describes a process in which an electron scatters from
a state in branch $\beta$ with momentum $k$ to a state in branch
$\alpha$ with momentum $k\!-\!q$ by emitting a phonon. In all these
processes, both total energy and momentum are conserved. The
emission and absorption rate coefficients are given by:
\begin{equation}
R^i_a(T_L) = \frac{N_q}{\tau_i} =\frac{1}{\tau_i} \frac{1}
{\exp(\hbar \omega /k_BT_L)-1}
\end{equation}
\begin{equation}
R^i_e(T_L) = \frac{N_{q}\!+1\!}{\tau_i} =R_a\exp(\hbar \omega /k_BT_L)
\end{equation}
where $T_L$ is the lattice temperature and $1/\tau_i$ stands for
the bare scattering rate for OP and ZB phonons that we assume to
be independent of carrier energy in a first approximation. In computing the
collision integral, we make the key ``ansatz'' that electrons and
lattice are in local thermal equilibrium (i.e. $T_L\! =\! T_L(x)\! =\!
T_{el}(x)$). We define the electron density as:
\begin{equation}
n_\alpha = \frac{1}{\pi} \int^{+\infty}_{-k(E_c)} f^\alpha(k) dk
\end{equation}
where $E_c$ is the bottom of the conduction band, and the current to be:
\begin{equation}
I =e (n_+-n_-) v_F
\end{equation}
where the $+(\!-\!)$ index corresponds to the branches with positive
(negative) Fermi velocity. Then integrating Eq.~\ref{boltzmann1}
over the momentum for each branch, and properly
accounting for different branches, we obtain:
\begin{equation}
v_F \partial_x \left(\!n^{\!+} \!-\! n^{\!-}\!\right) = 0 \label{chargecons}
\end{equation}
\begin{equation}
v_F \partial_{x\!}\! \left(n^{\!+}\!\! +\! n^{\!-}\!\right)\!
-\!\!\frac{2eF}{\pi \hbar}\! =\!\!2\! \!\int\!\! \!dk
C_{ph}\!(I,\!T(x)\!) \!\! = \!2
\tilde{C}_{ph}\!(I,\!T(x)\!)\label{boltzmann2}
\end{equation}
Eq.~\ref{chargecons} is the expression of the current conservation
in the system which by symmetry $\epsilon^\pm_{F1}\! =\!
\epsilon^\pm_{F2}$ and with charge neutrality in the CN yields
$\epsilon^+_{F(1,2)}\! =\!- \epsilon^-_{F(1,2)}$. Integrating 
Eq.~\ref{boltzmann2} over the length of the nanotube $L$, and assuming
equal electron densities at the contacts, we find the voltage drop
$V_{DS}$ along the nanotube given by:
\begin{equation}
V_{DS} = -\frac{\pi \hbar}{e}\int_{-L/2}^{L/2} dx\: \tilde{C}_{ph}(I,T(x))
\label{voltage}
\end{equation}
This equation implicitly depends on the current and the
temperature profile along the nanotube, and must be solved
self-consistently with the heat transport equation to obtain both
current and temperature profile.

As a particular case, it is interesting to compute the {\it IV}
relation from Eq.~\ref{voltage} by assuming only OP phonon
scattering at constant temperature (i.e. $T(x)\!=\!T_0$) in the CN.
The results are plotted in Fig.~\ref{iv_const_diag}(b) where the
high bias regime exhibits an asymptotic behavior independent of
temperature which is given by:
\begin{equation}
V_{DS}(I) = \frac{1}{G_0}\frac{L}{v_F\tau_{op}}(I - I_{\omega_{op}})
\label{currthres}
\end{equation}
where $I_{\omega_{op}}\!=\!e\omega_{op}/\pi$ is the threshold
current, corresponding to the onset of electron backscattering by
OP phonons \cite{yao} and $G_0\!=\!2e^2/h$ is the quantum conductance.
We point out that Eq.~\ref{currthres} is not consistent with the current interpretation
of electrons accelerated ballistically in the electric field
until acquiring enough energy to emit a phonon, but rather results
from the imbalance between the population of the energy branches
with positive and negative Fermi velocities. By considering ZB phonons, the
voltage drop in the m-SWNT is a linear combination of expressions
similar to Eq.~\ref{currthres}, which would still result in a
threshold current, but with a more complicated expression.

\begin{figure}[htbp]
  \leavevmode \centering
    \includegraphics[width=3in]{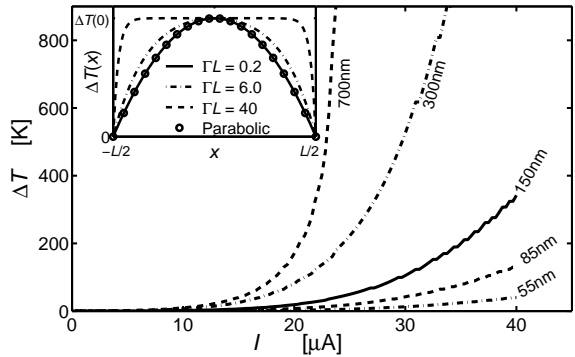}
    \vspace{-0.15in}
  \caption{\label{itchar} Temperature difference between the middle of
the tube and the leads as a function of the current for different CN lengths 
\cite{javey}. Inset: Temperature
profile along the
NT for different values of $\Gamma L$.}
\end{figure}

When heating effects become relevant, thermal dissipation is taken
into account self-consistently with Eq.~\ref{voltage}. We consider
that two mechanisms for the heat dissipation:
i) diffusion through the supporting substrate (if the CN
stands on one), and ii) flow through the contacts. Hence,
defining $\Delta T \!=\! T(x)\!\!-\!\!T_0$ (where $T_0$ is the temperature
of the substrate and leads), the heat equation becomes \cite{durkan}:
\begin{equation}
-\kappa \frac{d^2 \Delta T}{dx^2} + \gamma \Delta T = q^* \label{heateq}
\end{equation}
where $\kappa$ is the thermal conductivity, $\gamma$ is the
coupling coefficient with the substrate and $q^*$ is the power
dissipated per unit volume. Here we make the usual
approximation that process (i) is proportional to the local
temperature difference between CN and substrate. In
our calculations both the thermal conductivity and the coupling
coefficient are assumed to remain constant along the tube. The
coefficient $\gamma$ is given \cite{durkan} by:
\begin{equation}
\gamma = \frac{\kappa_{sub}}{t\, d}
\end{equation}
where $\kappa_{sub}$, $t$ and $d$ are the thermal conductivity of
the substrate, the diameter of the nanotube  and the thickness of
substrate, respectively. We also
assume that the power is homogeneously generated along the CN  and given by
Joule's law:
\begin{equation}
q^* = j\:F
\end{equation}
where $j = I/A$ is current density through the effective
cross section $A$ and the electric field $F$ is given by
$F=|V_{DS}/L|$. Then the solution for temperature profile is given
by:
\begin{equation} \Delta T(x) = \frac{q^*}{\gamma L S}
 \left[1 -\frac {\cosh(\Gamma x)}{\cosh( \Gamma L/2)}\right] \label{temprof}
\end{equation}
where $\Gamma = \sqrt{\gamma/\kappa}$.

Different scenarios can take place depending on the value $\Gamma
L$ (see the inset of Fig. \ref{itchar}). On the one hand, for
$\Gamma L\!\!\ll\!\!1$ diffusion through the substrate is
negligible and the temperature profile exhibits a parabolic shape.
On the other hand, for $\Gamma L\!\! \gg\!\!1$, heat basically
dissipates through the substrate and the temperature is almost
constant along the CN. This latter situation occurs in long tubes
strongly coupled to the substrate. In any case, the highest
temperature point is at the middle of the tube.

\begin{figure}[htbp]
  \leavevmode \centering
    \includegraphics[width=3in]{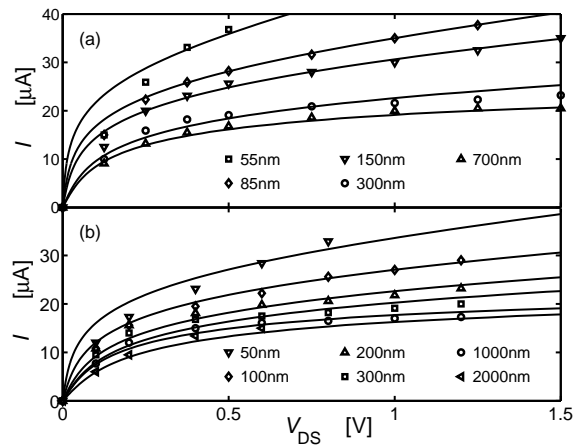}
  \vspace{-0.15in}
  \caption{\label{ivchar} Comparison between 
theoretical and experimental $IV$ characteristics 
for different CN lengths: (a)~
Ref.~\cite{javey} (b)~Ref.~\cite{park}.}
\end{figure}

In our calculations we use $T_0$=300K. The energies of OP and
ZB phonons are $\hbar \omega_{op}$=0.20eV and $\hbar
\omega_{zb}$=0.16eV, respectively. We use the standard accepted
value for the thermal conductivity $\kappa$ (30W/cmK) \cite{che} 
and $\Gamma \!=\! 10^{11}\mbox{W\,cm}^{-3}\mbox{K}^{-1}$.
Fig.~\ref{itchar} shows the temperature difference in the middle
of the tube ($\Delta T(0)$) as a function of $I$ for different CN
length, corresponding to the data of Ref. \cite{javey}. The longer
the nanotube, the faster the rise in temperature as the threshold
current is overcome. This is due to the fact that dissipation
occurs over a longer distance while heat removal mainly takes
place at the contacts in 1D structures. Estimates for the
breakdown temperature correspond to 800$^\circ$C \cite{radosav}.
Therefore, short tubes are expected to carry larger currents
before thermal breakdown. As shown in Fig.~\ref{ivchar}, the
results for the $IV$ characteristics are in good agreement with
the experimental data \cite{javey, park}. Deviations in the low
bias regime are mainly due to the absence of acoustic phonons
scattering in our model. For the sake of simplicity, we assume
relaxation times for OP and ZB phonons with equal values, which
are $\tau\!=\!(13\!\pm\! 2)$fs for the first \cite{javey} and
$\tau\! =\!(6.9\!\pm\!1.5)$fs for the second \cite{park} sets
of experimental data. The difference between the two values could
be due to the fact that CNs may have different diameters with
different phonons spectra (breathing modes) \cite{leroy} in each
case, and contact quality. Nevertheless, the obtained mean free
paths (between 6nm and 10nm) are consistent with experimental
previous estimates \cite{javey,park}. For long tubes (i.e.
$L\!=$700nm in the first case \cite{javey} and for
$L\!\ge\!$1000nm in the second case \cite{park}), we need to
increase in both cases the relaxation times to $(27\!\pm\!3)$fs to
fit the experimental data. Since dissipation is considerably
stronger in long CNs, these longer times may be associated to the
emergence of non-linear thermal effects in m-SWNT thermal
conductivity not taken into account in Eq.~\ref{heateq}, and for
which the temperature and geometry dependence at room temperature
or higher remains an open issue~\cite{cao,yamamoto}.

\begin{figure}[htbp]
  \leavevmode \centering
    \includegraphics[width=3in]{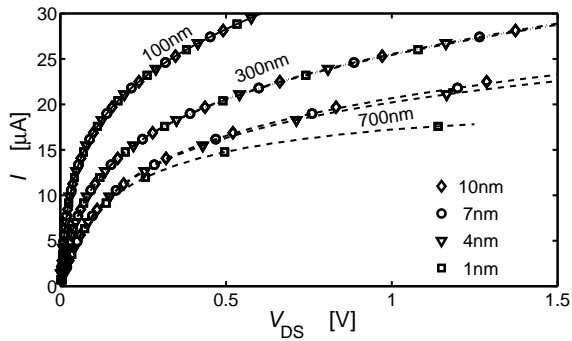}
    \vspace{-0.15in}
  \caption{\label{diamdep} $IV$ characteristics for different tube
lengths and diameters. Short tubes (100nm and 300nm) are
practically independent of the diameter, while in the long tube
(700nm),
deviations appear for the small diameter (1nm).}
\end{figure}
Finally, Fig.~\ref{diamdep} shows the {\it IV}-characteristics obtained for
CNs of different diameters and lengths, assuming that
both the thermal conductivity and the relaxation time are equal
among tubes. Despite this strong assumption, the relevant issue
to emphasize here is the weak dependence of the $IV$
characteristics on the size of the CN. Appreciable deviations can
be observed only in 700nm tube (dashed line) for small diameters
($\sim$1nm). This weak relation is consistent with the
interpretation of Collins et al.~\cite{collins1,collins2} for the
electrical breakdown under electrical stress observed in MWNTs,
where different layers in a
MWNT (separated by about 0.4nm) carry similar currents in the high
bias regime. The breakdown of successive carbon layers produces 
approximately constant diminutions of the current in the high bias
regime because  {\it IV} characteristics are geometry independent. 
Moreover, the highest temperature arises at the CN mid-length 
(inset of Fig.~\ref{itchar})
and therefore, electrical breakdown is,
as experimentally observed, expected to take place there too.

In conclusions, we have shown that the consideration of a
thermalized electron distribution in local equilibrium
with a non-homogenously heated lattice through OP and ZB 
scattering determined
self-consistently by the current level account
for the non-linear $IV$ characteristics of the m-SWNTs in the high
bias regime. The magnitude of the temperature variation
as a function of the CN lengths is consistent with the
occurrence of thermal breakdown at mid-length for long
CN under electrical stress. While the dependence of thermal conductivity 
on temperature still reamins under investigation, our self-consistent 
model provides a coherent picture of the onset of thermal effects 
with electronic transport in m-SWNT.

\begin{acknowledgments}
The authors are indebted to A. Cangellaris for fruitful discussion.
This work was supported by the Beckman Institute for Advance
Science and Technology and NSF - Network of Computational
Nanotechnology.
\end{acknowledgments}

\begin{thebibliography}{99}
\bibitem{saito_dres} R. Saito, G. Dresselhaus, and M. Dresselhaus, {\it ``Physical
Properties of Carbon Nanotubes,''} Imperial College Press (1998).
\bibitem{mceuen} P.L. McEuen, M.S. Fuhrer, Hongkun Park, IEEE Transactions on Nanotechnology, {\bf 1}, 78 (2002).
\bibitem{avouris} Ph. Avouris, J. Appenzeller, R. Martel, S. Wind, Proceedings of
the IEEE, {\bf 91}, (1772) 2003.
\bibitem{yao} Z. Yao, C. L. Kane, and C. Dekker, Phys. Rev. Lett. {\bf 84}, 2941 (2000).
\bibitem{javey} A. Javey, J. Guo, M. Paulsson, Q. Wang, D. Mann, M. Lundstrom, and H.
Dai, Phys. Rev. Lett. {\bf 92}, 106804 (2004).
\bibitem{park} J.Y. Park, S. Rosenblatt, Y. Yaish, V.
Sazonova, H. \"Ust\"unel, S. Braig, T.A. Arias, P.W. Brouwer, P.L.
McEuen, Nano Lett. {\bf 4}, 517 (2004).
\bibitem{collins1} P.G. Collins, M. Hersam, M. Arnold, R. Martel, and Ph. Avouris,
Phyis. Rev. Lett. {\bf 86}, 3128 (2001).
\bibitem{collins2} P.G. Collins, M.S. Arnold, P. Avouris, Science
{\bf 292}, 706 (2001).
\bibitem{muller} R.S. Muller, T.I. Kamins, {\it `` Device Electronics for Integrated Circuits''},
3rd. Edition, John Wiley \& Sons (2002).
\bibitem{jishi} R.A. Jishi, D. Inomata, K. Nakao, M.S. Dresselhaus, G. Dresselhaus,
J. Phys. Soc. Japan, {\bf 63}, 2252 (1994).
\bibitem{saito} R. Saito, T. Takeya, T. Kimura, G. Dresselhaus and M.S. Dresselhaus,
Phys. Rev. B {\bf 57}, 4145 (1998).
\bibitem{perebeinos} V. Perebeinos, J. Tersoff, and Ph. Avouris, Phys. Rev.
Lett. {\bf 94}, 027402 (2005).
\bibitem{durkan} C. Durkan, M.A. Schneider, and M.E. Welland, J. Appl. Phys. {\bf 86}, 1280 (1999).
\bibitem{che}
J. Che, T. Cagin, and W.A. Goddard III, Nanotechnology, {\bf 11},
65 (2000).
\bibitem{radosav} M. Radosavljevic, J. Lefebvre, and A. T.
Johnson, Phys. Rev. B {\bf 64}, 241307(R) (2001).
\bibitem{leroy} B.J. LeRoy, S.G. Lemay, J. Kong and C. Dekker, Nature {\bf 432}, 371 (2004).
\bibitem{yamamoto} T. Yamamoto, S. Watanabe, and K. Watanabe, Phys. Rev.
Lett. {\bf 92}, 075502 (2004).
\bibitem{cao} J.X. Cao, X.H. Yan, Y. Xiao, and J.W. Ding, Phys. Rev. B {\bf 69}, 073407 (2004).



\end{thebibliography}

\end{document}